# Inkjet Printed Liquid Crystal Droplet for Complex Beam Manipulation

*Mengmeng Li †, Chao He †, Steve J. Elston, Yifei Ma, Bohan Chen, Zimo Zhao, Xuke Qiu, Alfonso A. Castrejón-Pita, and Stephen M. Morris\**

*Department of Engineering Science, University of Oxford, Parks Road, Oxford, OX1 3PJ, UK*

*E-mail: Mengmeng.li@eng.ox.ac.uk ; Chao.he@eng.ox.ac.uk; stephen.morris@eng.ox.ac.uk*

**Abstract**

The inkjet-fabricated liquid crystal (LC) droplet device not only capitalizes on the intrinsic birefringence properties of liquid crystals but also leverages the hemispherical shape of droplet devices on substrates. This configuration facilitates self-alignment of the LC director under the influence of surface tension. The LC droplet devices we fabricated are capable of intricate beam manipulation, encompassing both generation and analysis of light beams. Such devices possess substantial prospective applications in the fields of optical communications and light beam characterization, highlighting their significant potential for advancement in optical technologies.

**Introduction**

Inkjet printing is a crucial technology utilized across various fields, including biosensing[1], flexible electronics[2, 3], wearable electronic devices[4, 5], sensor technology[6, 7], OLED display technology[8, 9], medical devices[10], biomedical applications[11], and MEMS devices[12-14], among others. This inkjet printing technology streamlines manufacturing, reducing costs and time by allowing precise, customizable, and scalable production on various substrates, with the added benefits of energy efficiency and minimal waste. The utilization of inkjet printing for depositing liquid crystal (LC) mixtures offers an alternative fabrication technique in the field of optical components, effectively harnessing the distinctive birefringence properties inherent to LC materials. Recent studies have explored the application of printed LC droplet devices in the domain of lens technology, specifically in the development of tuneable[15] and bifocal microlens arrays[16]. This advancement is facilitated through the application of electric fields or thermal modulation. Nonetheless, there exists potential for further exploitation of the birefringence properties inherent to liquid crystals, particularly in the generation of light beams.

In this study, we have broadened the application of inkjet printing technology in the LC optoelectronic devices, focusing on both beam generating and analysing using inkjet-printed LC droplets. We first utilize printed LC droplets with various diameters and employ different combination of thin film based half-wave and quarter-wave plates to produce an array of optical skyrmionium types. Our observations indicate that the size of the printed LC droplets and the varying arrangements of optical components influence the resulting types of optical skyrmions. This finding implies that by precisely adjusting the droplet size and modifying the optical component configurations, it's feasible to generate customized skyrmion structures, each uniquely suited for different applications or specialized research needs. Secondly, our investigation extends to the precise measurement of polarized light beams, showcasing the utility of printed LC droplets as a Stokes polarimeter. By capitalizing on the intrinsic birefringence of the droplets, which acts as infinite polarization analyzing channels, we demonstrate the significant potential of this approach for advanced optical analysis. These droplets enable single-shot measurements with an error margin of less than 1% in the determination of Stokes parameters. This accuracy highlights the effectiveness of the printed LC device in analysing the polarization of unknown light sources. The use of inkjet printing provides a cost-effective and straightforward approach to fabricating LC-based optical components, enabling the rapid production of devices with specific functionalities. This technique facilitates the creation of structures, such as those required for being a multifunctional optical manipulator, by allowing precise control over the placement and patterning of liquid crystal materials. The resulting printed LC devices exhibit outstanding functionality, making them highly versatile in a wide array of applications, including optical communication and biotissue measurement. This versatility not only demonstrates the technological innovation of our approach but also underscores the potential for its widespread application in various scientific and industrial fields.

**Results**

Fabrication procedure

The device fabrication process is characterized by its simplicity and convenience, making it suitable for efficient and reliable production as shown in **Figure 1**. To achieve this, LC droplets are precisely printed onto glass slides that have undergone treatment with a lecithin solution as shown in Figure 1 i). This treatment ensures the establishment of a homotropic alignment layer, crucial for the desired alignment of the LC droplets. The LC director aligns perpendicular both to the substrate and the interface between the air and the liquid crystal. The fabrication process is seamlessly controlled by a computer, offering a high degree of automation. The electric signal (Figure 1 ii)) was generated by the computer and then sent to the piezoelectric actuator inside the printhead (Figure 1 iii)). This automated approach enhances the accuracy and reproducibility of the printing process. By controlling the dwell and echo time of the input signal, the different printing process could be achieved. Figure 1 iv) shows the shadowgraph images capturing the droplet formation process in the air. This visual representation serves to document the successful implementation of the fabrication method and demonstrates the capability of the printing technique to generate well-defined and organized LC droplet structures. The droplet diameter during mid-air suspension reaches a minimum of approximately 70 μm due to the 80 μm inner diameter of the nozzle. Upon contact with the substrate, it expands to around 100 μm. To fabricate larger droplets, a method involving the deposition of 62 droplets at a precise location was employed, resulting in a single droplet with a final diameter of 420 μm. Figure 1 v) shows the side view of the printed LC droplet which is hemispherical in shape on the glass substrate. And the laser beam goes through the printed LC droplet device as Figure 1 vi) shows when the device is working as Optical Skyrmion Generator and Stokes polarimeter with the specific LC director distribution.

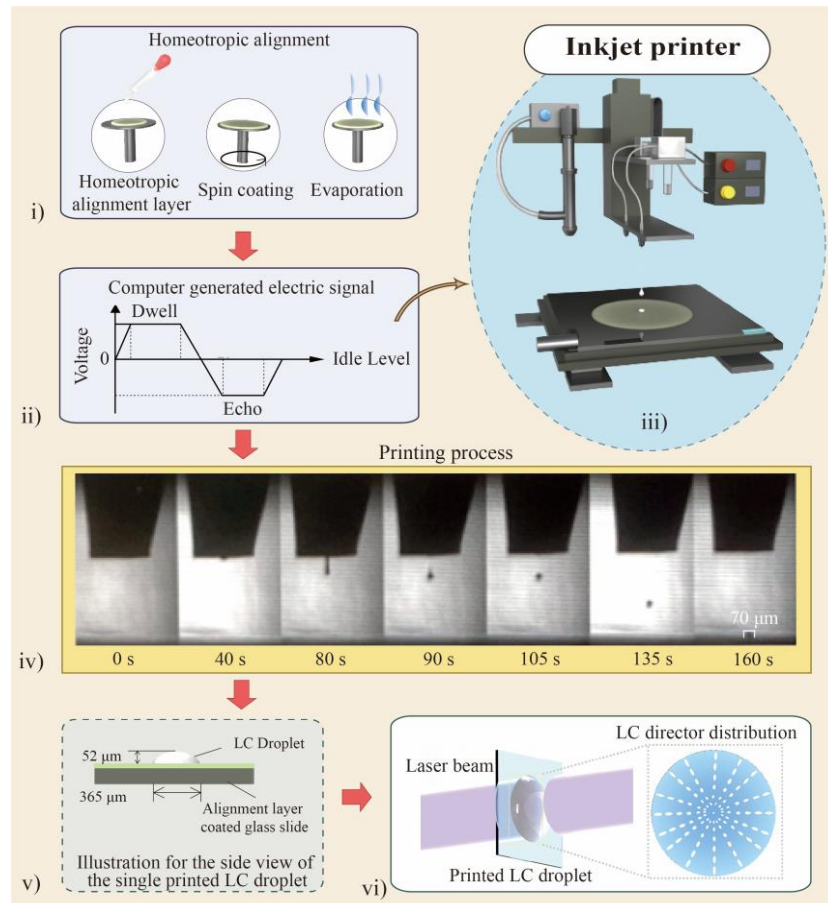

Figure 1. Manufacturing process of the LC droplet device. i) The spin coating process with a homeotropic layer. ii) Voltage waveform used to drive the piezoelectric transducer in the printhead of the iii) Microfab inkjet printer. iv) Shadowgraph images of the printing process showing the formation of the droplet in air. v) Side image of the printed LC droplet which forms a hemispherical shape. vi) Illustration of the illumination of the printed LC device and the corresponding LC director distribution inside the droplet. The printing process employs a nozzle with an 80 μm inner diameter. The droplet diameter in mid-air approximates to 70 μm, extending to 100 μm upon substrate contact. To engineer a larger droplet, a strategy involving multi-droplet deposition at a single point was utilized, yielding a droplet with a footprint diameter of 420 μm.

# Experimental results:

**Optical Skyrmion Generator**

Skyrmions, first explored by Tony Skyrme in nuclear physics[17], have garnered significant attention in condensed matter physics for their stability and energy efficiency, offering promising applications in spintronics and magnetic data storage. Their expansion into photonics, where they manifest as confined wave packets in optical materials, opens avenues for understanding and applications in fields like optical tweezers[18, 19], microscopy[20], and communications[21]. Concurrently, the study of optical topological quasiparticles, including merons[22, 23] and hopfions[24-26], is advancing, driven by their unique topological properties and resilience. These developments could revolutionize optical data storage and processing[27], leveraging their complex structures for enhanced data density and revealing new phases of matter for advanced photonic device development.

In this study, we developed a LC droplet device capable of generating optical skyrmions, as evidenced by our experiments and depicted in Figure 2 a) i). By employing printed LC droplets, our apparatus not only produces higher-order optical skyrmions but also complex multi-bimerons and hybrid states, with comparative analysis and hue-coded visual representations for clarity. We demonstrate the device's capability to create high-order skyrmionium as shown in Figure 2 b) i), closely matching the theoretical simulations in Figure 2 b) ii), thus validating our approach against theoretical models. Additionally, we investigate the generation of bimeronium as shown in Figure 2 b) iii), consisting of four sub-bimerons with specific vector orientations, which aligns well with simulations in Figure 2 b) iv), further confirming the device's efficacy in generating advanced topological entities. This work establishes the LC droplet device as an effective generator of optical skyrmions, showcasing its potential in accurately replicating complex topological structures.

By changing the different states of polarization (SOPs) as input, the optical skyrmion and skyrmion field generated after passing through the device also differs. And the results vary according to the diameter of the printed LC droplets. **Figure 2 c)** shows the results. The first line demonstrated the different incident SOPs from the left-handed circular polarised light, linear polarised light, and right-handed circular polarised light. The generated results showed in the Figure 2 c) i) α and β is what discussed before in Figure 2 b). Upon integrating the thin film droplet device with a half-wave plate and a quarter-wave plate, diverse optical skyrmions can be generated contingent upon the incident SOPs. This posits that our LC thin film-based droplet device holds promise as a generator for optical skyrmions. Notwithstanding, all generated optical skyrmions consistently exhibit a flux value of zero, which presents challenges for its immediate application in optical communications. There is a pressing need for further investigation to adapt it effectively for communication applications. Nonetheless, this serves as a testament to the latent potential of the LC thin film-based droplet device in producing optical skyrmions.

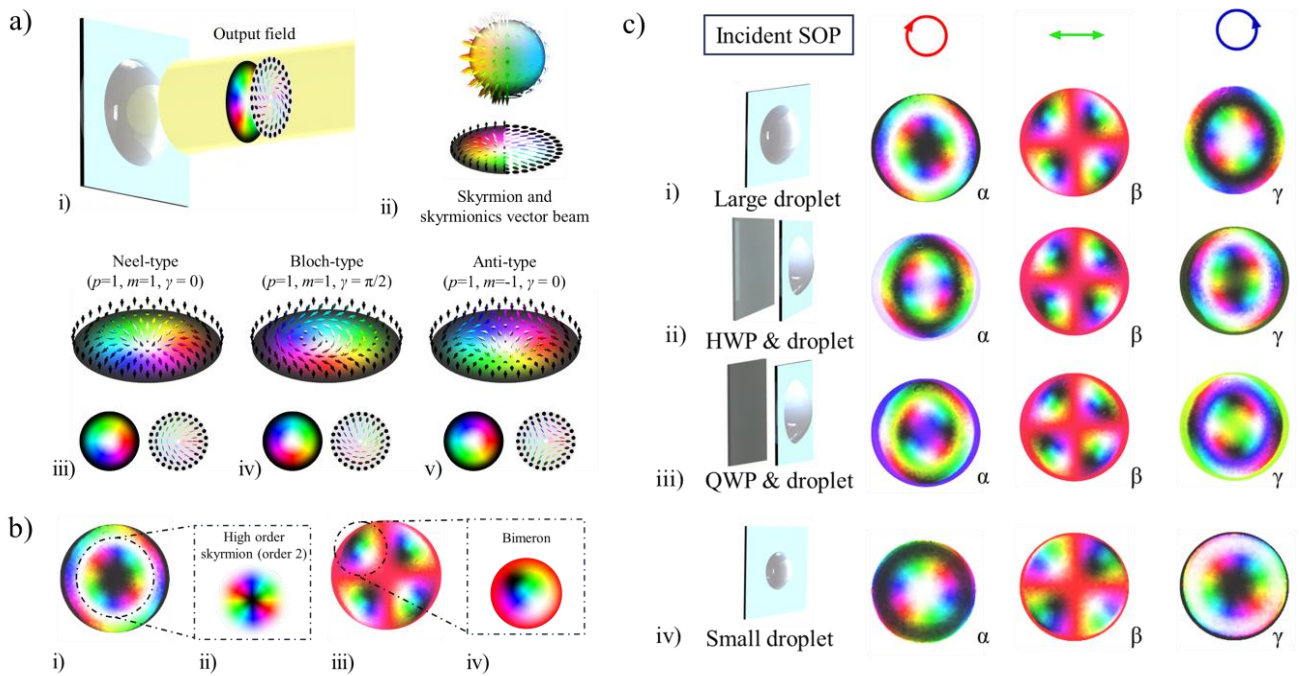

Figure 2. Generation of Optical Skyrmions via the Printed LC Droplet Device. a) Optical Skyrmion Field and Varieties: i) Display of the optical skyrmion field featuring various polarization states. ii) Vector distribution within an optical skyrmion. iii) Néel-type skyrmion. iv) Bloch-type skyrmion. v) Anti-type skyrmion. Each type of the skyrmion was characterized by specific polarity (p), vorticity (m), and helicity (γ). b)

Skyrmionium Comparison: i) Experimentally produced skyrmionium with left-handed circularly polarized light input SOP. ii) Simulation of the internal structure of i). iii) Skyrmionium generated experimentally with linearly polarized light input SOP. iv) Simulation depicting a quarter of the structure in (iii), corresponding to the experimental setup. c) Skyrmion Beams via Varied SOPs and Device Configurations: i) Skyrmionium produced with a printed LC droplet device featuring a 420 μm diameter. ii) Skyrmionium generated using a half-wave plate combined with a printed LC droplet device of 420 μm diameter. iii) Creation of skyrmionium through a quarter-wave plate and a printed LC droplet device, also with a 420 μm diameter. iv) Skyrmionium formed with a printed LC droplet device, utilizing a smaller droplet diameter of 260 μm. Each setup demonstrates the impact of different input SOPs - α for left-handed circular polarized light, β for linear polarized light, and γ for right-handed circular polarized light - on the formation and characteristics of optical skyrmions.

**Optical Polarisation Analyser**

Besides its application in the generation of complex structured light, the printed LC device also finds utility in the measurement of the polarization state of the light beam. Capitalizing on the birefringence properties of LC droplets, these devices are well-suited for use in polarization analysing channels. This makes them integral to Stokes polarimeters, which are crucial for the accurate characterization of light's polarization state. Stokes polarimeter is a robust scientific approach for scrutinizing the polarization properties of light, providing critical information pertinent to the interactions between light and matter. This technique bolsters progress across myriad domains, including biomedicine[28, 29], materials science[30, 31], and environmental studies[32, 33]. Stokes polarimeters are principally divided into two categories: time-sequential and simultaneous data acquisition instruments. The time-sequential devices have the advantage of being cost-effective and compatible with a variety of detectors but are generally slower in operation, capturing one parameter at a time. In contrast, simultaneous polarimeters amplify temporal resolution and expedite the data collection process. However, ongoing research focuses on enhancing and expanding the capabilities of Stokes polarimetry in various scientific and technological fields. In our research, we primarily utilize the Division of wavefront method[34]. These LC droplets, functioning as lenses, exhibit radial refractive index variability and serve as a rotating waveplate in the azimuth direction, while undergoing birefringence modifications in the radial direction, thereby facilitating simultaneous measurement of the Stokes parameters. This instrument, owing to its compact and sleek design, can be assimilated into various optical systems without compromising the accuracy of measurements.

To enable this capability, the LC droplet device incorporates an imaging lens, a polarizer, and a CCD camera, as depicted in Figure 3 b). This configuration is designed to capture detailed information on incident light beams. The intrinsic birefringence of the LC directors modulates the propagation path of light within the droplet. This modulation varies across the droplet's different regions due to disparities in the LC director distribution, thereby affecting the incident light in varying magnitudes. As a result, when a uniform light beam is processed through the LC droplet-based Stokes polarimeter, its interaction with the device varies spatially, leading to differential effects at distinct locations. The unique optical properties of liquid crystals (LC) can be further analysed using a Mueller matrix polarimeter, details of which are provided in the supplementary information. This analysis allows us to uncover detailed information, including the retardance and the distribution of the fast axis throughout the LC droplet. As previously mentioned, only four different modulation channels are necessary to measure the four unknown parameters of the Stokes vector. However, the printed droplet device offers more than four different channels, which could lead to a reduction in measurement error, thereby increasing the accuracy of the Stokes vector determination. Figure 3 c) depicts the experimental setup capable of generating varied SOPs through a polarization state generator (PSG). However, as indicated in the supplementary information, in theory, only four distinct points on the printed LC droplet are required for a comprehensive analysis, given that the Stokes parameters consist of four components. The selection of these four points greatly influences the measurement accuracy, making it imperative to choose them with care to minimize errors. Prior studies on optimizing the Stokes polarimeter have tackled this challenge by identifying four optimal points based on criteria that the fast axis orientations should be 15.12°, 51.69°, 128.31°, and 164.88°, with the retardance at these points being 132°[35]. This strategy ensures minimized measurement errors and enhances the precision of polarization analysis.

The foundational principles of the Stokes polarimeter's operation, as previously described, are essential for its application. The experimental protocol utilizes the Mueller matrix polarimeter[36] to determine the optical properties of the device, such as retardance and the distribution of the fast axis. This protocol is structured into two essential phases: calibration and measurement. Initially, the calibration phase involves computing the instrument matrix $A$ for the printed LC droplet Stokes polarimeter, according to Equation (1). This step is crucial for establishing a precise correlation between the measured Stokes parameters and the actual SOP of the incoming light. Subsequently, the measurement phase applies the calculated matrix $A$ to identify the Stokes parameters of the input SOP, as specified by Equation (2). An explanation of the calculation process, vital for the calibration and measurement stages, is detailed in the supplementary information. This thorough documentation provides clear instructions and insights into the experimental procedures necessary for the effective operation of the polarimeter.

$$A = I \cdot S^{-1} \tag{1}$$

$$S = A^{-1} \cdot I \tag{2}$$

As described before, the four specific points on the printed LC droplet device were selected in the optimization process. These droplets need to satisfy the condition that the fast axis is 15.12°, 51.69°, 128.31°, and 164.88° and in the meantime the retardance need to be 132° as shown in the lower left corner of **Figure 3** d) i). While there are more than 4 points that could satisfy the condition, in fact there are four series of the points could be used. Therefor the second method is to use these four series points to calculate the instrumental matrix and the input beam's SOP as shown in the lower left corner of Figure 3 d) ii). Nevertheless, more points used in the calculation may help to reduce the errors regardless of the efficiency. The third method is to use all the points on the printed LC droplet in the whole process as shown in the same place of Figure 3 d) iii). In each row of Figure 3 d), the left panel illustrates a polarizer rotating in 10° increments which could generate linear polarised light, and the right panel shows the polarizer fixed with the quarter-wave plate rotating which could generate the circular polarised light. On the top of Figure 3 d) shows the different polarization state of the input light. The lines depicted in the figure illustrate the results obtained through simulation, whereas the points denote the outcomes of the experiments. Within the framework of the data processing paradigm, the parameter denoted as $S_0$, encapsulating the total intensity of the light, has been normalised to unity. Concurrently, Figure 3 d) presents the processed results pertaining to the remaining triad of Stokes parameters. Specifically, the results for $S_1$ are represented by the red lines or points, those for $S_2$ are indicated by the blue elements, and the green ones correspond to the results for $S_3$. The simulation and the experimental results demonstrated a strong match whether using the linear polarised light or the circular polarised light as the input.

In order to assess the relative performance of the three methodologies employed in our measurement process, a comprehensive error analysis was conducted focusing on the three Stokes parameters. This analytical examination is thoroughly documented in the supplementary information. It was observed that employing the entire array of points on the printed LC droplet device yielded the highest accuracy in measurements. This conclusion, however, is drawn without taking the time factor into consideration, as this approach is more time-intensive compared to the other methods evaluated.

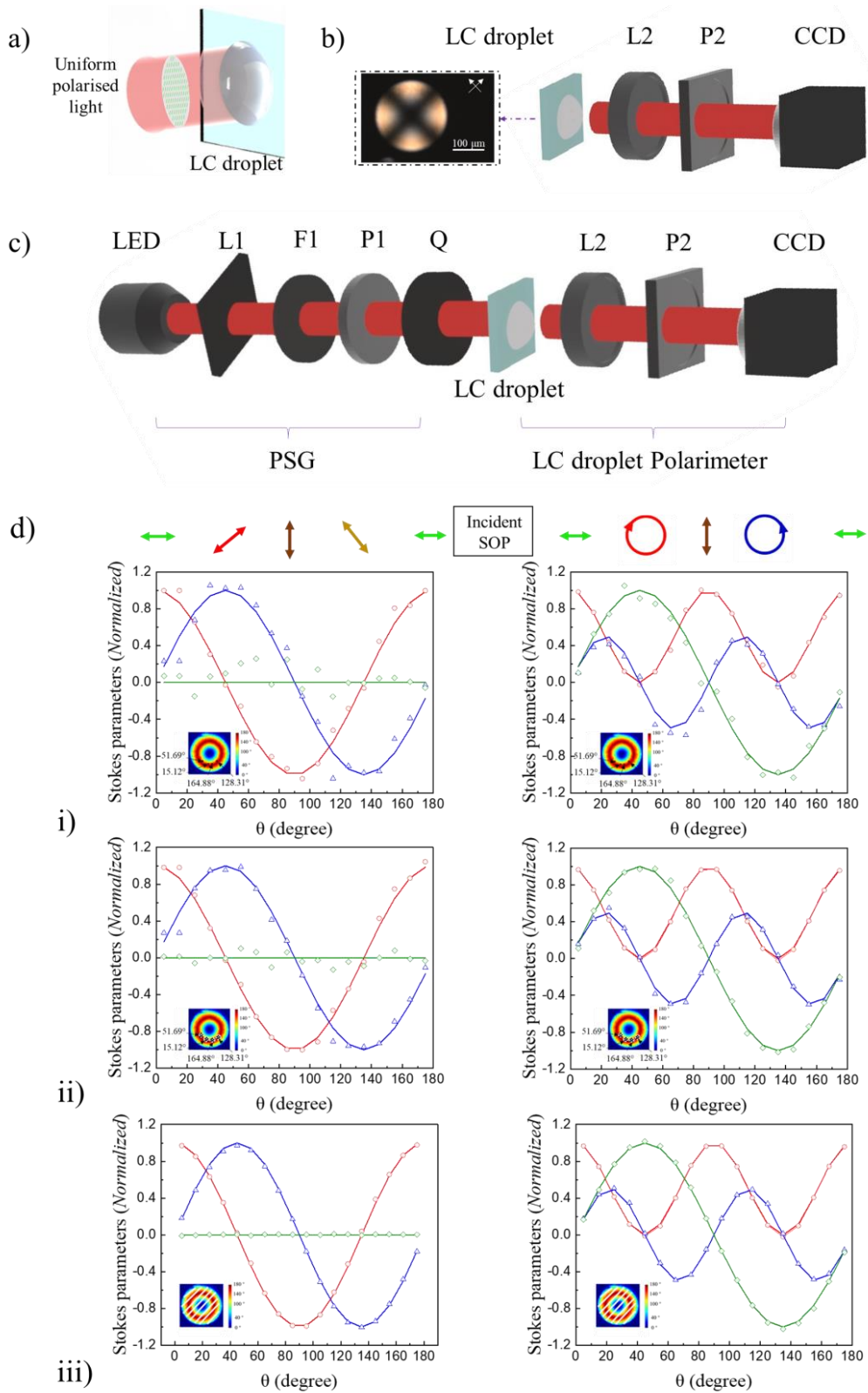

Figure 3. Deployment and the measurement results of the printed LC droplet in a Stokes polarimeter. a) Demonstration of the passage of uniform light that goes through the LC droplet. b) Configuration of the printed LC droplet Stokes polarimeter (a polarising optical microscope image of the printed droplet is also shown in the dashed lines – the white single-headed arrows correspond to the orientations of the polariser and analyser). c) Experimental assembly of the whole system including a polarization state generator (PSG) for the generation of different states of polarization of light and the LC droplet Stokes polarimeter. The diameter of the printed LC droplet is 420 μm. L1 is collimating lens; F is the band pass filter; P1 and P2 are the polarisers; Q is a quarter wave plate; L2 is an imaging lens. d) Results of different polarisation states measured by the printed LC droplet Stokes polarimeter. Measurements were obtained by selecting different points used in the measurement process located on the printed LC droplet (as shown by the inset in each plot. i) Four selected points that meet a condition of 132° retardance. Each of these points has its fast axis oriented at one of the following angles: 15.12°, 51.69°, 128.31°, and 164.88°. ii) Four groups of points that all meet the 132° retardance condition. Each group is categorized based on the fast axis orientation, corresponding to one of these angles: 15.12°, 51.69°, 128.31°, and 164.88°. iii) All points on the LC droplet device are used for the measurement process. The left-hand plots show experimental results for $S_1$ (open circles), $S_2$ (open triangles), and $S_3$ (open squares). Simulated results are displayed as red, blue, and green lines for $S_1$, $S_2$, and $S_3$, respectively, and these correspond to the Stokes parameters measured for various incident States of Polarization (SOPs). To obtain different incident SOPs, the polarizer in the Polarization State Generator (PSG) was rotated in 18 evenly divided steps across 180°. On the right, the Stokes parameters, in relation to different incident SOPs are illustrated. The PSG was adjusted with a fixed 0° polarizer and the quarter-wave plate was rotated in 18 equidistant steps through 180° to achieve a range of incident SOPs.

Discussion:

In this work, we have developed a novel thin film device using inkjet-printed liquid crystal (LC) droplets, serving dual functions as an optical Skyrmion generator and a Stokes polarimeter. Our approach to generating optical skyrmions, distinct from traditional methods, involves varying LC droplet dimensions and applying different input SOP with specific optical components like HWP and QWP, resulting in a range of Optical Skyrmionium structures. However, the topological numbers of these skyrmions consistently register as zero, highlighting potential limitations for their use in optical communication systems. This revelation, while presenting challenges, also lays the groundwork for future research endeavors. The application of technologies such as direct laser writing to alter the outer ring structure of the printed LC droplet may provide a viable solution to these limitations.

The Stokes polarimeter aspect of the device, assessed through rigorous experiments, demonstrates high accuracy in measuring incident SOPs, with a measurement error of about 1% or lower. The advantages of using inkjet printing technology include streamlined, cost-effective fabrication and the ability to customize the LC droplet size, ranging from 1 μm to several hundred microns. This customization, leveraging the birefringence properties of liquid crystals, enables single-shot measurements and streamlines the measurement system by obviating the need for mechanical adjustments. However, the surface homotropic treatment of the printed LC droplet may encounter decay issues. This problem could potentially be resolved by employing improved encapsulation techniques within a vacuum.

The dual functionality of our device, as both an optical Skyrmion generator and a Stokes polarimeter, is significant. It holds promise for advancing optical communication applications, particularly in generating high-order optical skyrmionium and bimeronium. The precision and simplicity of the Stokes polarimeter portion underscore its potential in intricate polarization measurements, enhancing the reliability and robustness of outcomes in various applications. This multifunctional device, simple in construction yet effective, represents a substantial advancement for information transmission against complex backgrounds and for nuanced detection of polarization signals, offering exciting prospects for future developments in optical technologies.